\documentclass[twocolumn,preprintnumbers,secnumarabic, nobibnotes, aps, showpacs, pra]{revtex4}
\usepackage{amssymb}
\usepackage{amsmath}
\usepackage{graphicx}
\usepackage{dcolumn}
\usepackage{bm}
\usepackage{braket}
\usepackage{mathrsfs}
\begin{document}
\title{Coherence Enhanced Transient Lasing in XUV Regime}
\author{P. K. Jha\footnote{Email: pkjha@physics.tamu.edu}, A. A. Svidzinsky, and M. O. Scully}
\affiliation{Texas A$\&$M University, College Station, Texas 77843 USA \\
 Princeton University, Princeton, New Jersey 08544, USA}
\date{\today}
\begin{abstract}
We report the effect of a coherent drive on transient lasing in three-level $\Lambda $ and $\Xi$ configurations ($c\leftrightarrow a\leftrightarrow b$). We show that the presence of a resonant coherent drive on the $a\leftrightarrow c$ optical transition can yield an order of magnitude enhancement of the output laser energy on a $a\rightarrow b$ XUV or X-ray transition than with no coherent drive. We demonstrate the crucial role of coherence $\varrho _{ac}$ for the laser power enhancement. Contrary to the forward direction (with respect to the pump), where forward gain can be enhanced for some choice of the drive Rabi frequency $\Omega_{c}$, coherent drive on the $ac$ transition always suppresses the backward gain.
\end{abstract}
\maketitle
\section{Introduction}
There is currently interest in developing XUV and X-ray coherent sources~\cite{Suck90,Skin,S1} which are useful tool for high resolution microscopy of biological elements~\cite{Skin90}, crystallography and condensed matter in general. There are several methods for producing extreme ultra-violet lasing: for example, plasma-based recombination lasers~\cite{Suck85,Suck86} or using a capillary discharge \cite{Rocc94}, a free-electron laser \cite{Milt01}, optical field ionization of a gas cell \cite{Lemo95}. Coherent XUV radiation can also be produced by the generation of harmonics of an optical laser in a gas or plasma medium~\cite{Krausz,Sansone,Schultze}. 

\begin{figure}[b]
  \includegraphics[height=5.0cm,width=0.49\textwidth,angle=0]{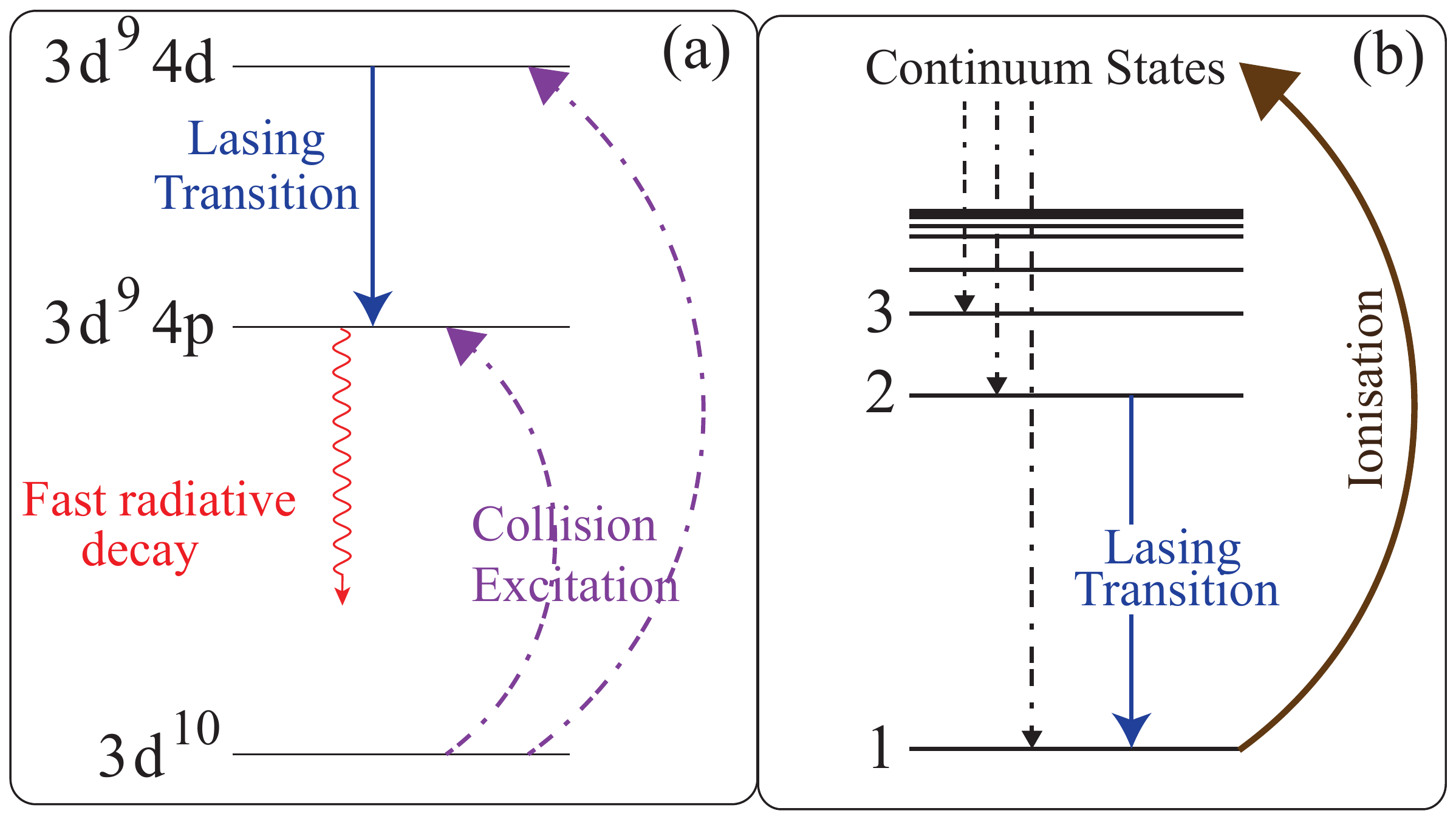}
  \caption{(a) Lasing in Ni-like ions. (b) Lasing in H-like ions}
  \label{Fig1}
\end{figure}

The quest for compact \textquotedblleft table-top\textquotedblright\ XUV and X-ray laser sources that can be used in individual research laboratories has motivated exploration of various excitation mechanisms, e.g., collisional~\cite{Matt85,Ros85}, recombinational~\cite{Suck85,Suck86,S1}, etc. Ionization-recombination excitation technique holds promise for making efficient lasers at shorter wavelengths and has been successfully implemented~\cite{S2}. In particular, a portable X-ray laser utilizing such excitation mechanism and operating in transient regime at $13.5$ nm has been demonstrated by Princeton group \cite{Koro96,Krus96,Golt99}. The laser uses H-like Li ions (see Fig. \ref{Fig1}b) as an active medium which are excited by ionization-recombination process. The basic idea of recombination lasers is that atoms are stripped off electrons in the initial step and then ions recombine by a three-body non-radiative recombination process which requires high density of electrons and prepares atoms or ions in highly excited states. By collisional de-excitation the population is transferred to lower excited states on a time scale of a few pico-seconds. For proper density, population inversion can be achieved on the probe-transition on a time scale of $10-100$ ps. We call this \textquotedblleft Inversion-Window\textquotedblright .

Apart from collisional recombination, schemes involving electron impact collisions were proposed to create inversion in Ne-like ions \cite{Zher76,Vino77,Vain78,Vino80}. Such schemes were later used for Ni-like ions
(see Fig. \ref{Fig1}$a$). Here the lasing transition is $3d^{9}4d\rightarrow 3d^{9}4p$. The two lasing levels are populated by electron collisions. While the radiative decay from the upper lasing level ($3d^{9}4d$) to the ground state ($3d^{10}$) is dipole forbidden, the fast radiative decay from the lower lasing level to the ground state makes it possible to achieve population inversion on the lasing transition and yield lasing in the \textquotedblleft Inversion-Window\textquotedblright. 
\begin{figure*}[t]
\centerline{\includegraphics[height=6.3cm,width=0.95\textwidth,angle=0]{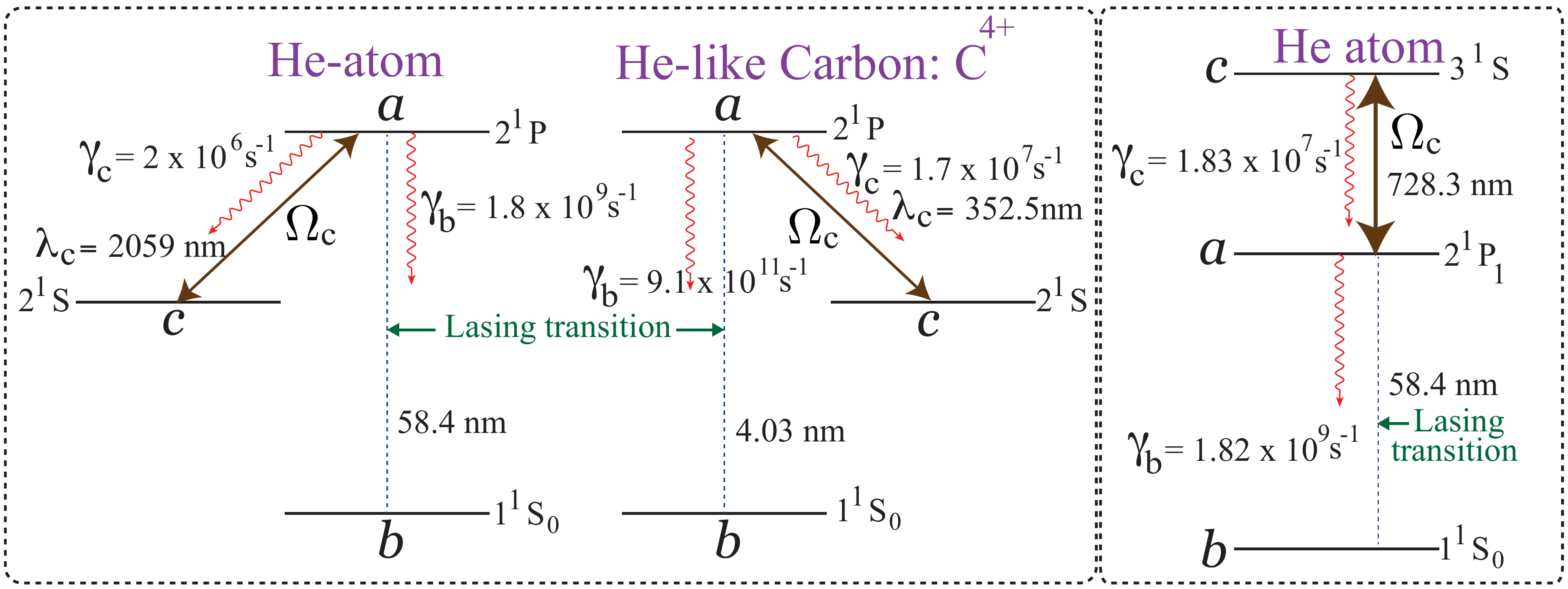}}
\caption{Energy level diagram of He atom and He-like Carbon in $\Lambda$ configuration (left box) and in He atom cascade configuration (right box). }
\label{Fig5}
\end{figure*}

In 1905 Einstein showed that the entropy of light displayed both wave and particle aspects which led him to introduce the concept of a photon \cite{Ein05}. Later in 1917 he discovered stimulated emission by using detailed
balance \cite{Ein17} and considering a beam of two-level atoms with ground state $b$ and excited state $a$ ($E_{a}-E_{b}=\hbar \omega $) interacting with electromagnetic field. Assuming that atomic populations in the excited $N_{a}$ and the ground $N_{b}$ states satisfy the rate equations \cite{Scul74}
\begin{equation}
\dot{N}_{a}=-AN_{a}-BU(\omega )(N_{a}-N_{b}),
\end{equation}
\begin{equation}
\dot{N}_{b}=AN_{a}+BU(\omega )(N_{a}-N_{b}),
\end{equation}
where $AN_{a}$ is the rate of spontaneous emission and $BU(\omega)(N_{a}-N_{b})$ is the corresponding rate of stimulated process, we obtain that in equilibrium
\begin{equation}
\lbrack A+BU(\omega )]N_{a}=BU(\omega )N_{b}.  \label{eq:DB}
\end{equation}
This condition is referred to as detailed balancing. In equilibrium at temperature $T$ relation between atomic populations is given by the Boltzmann distribution
\begin{equation}
\frac{N_{a}}{N_{b}}=\exp (-\hbar \omega /k_{B}T).  \label{eq:BD}
\end{equation}
Combining Eqs. (\ref{eq:DB}) and (\ref{eq:BD}) and using the Planck formula for the photon energy density per unit frequency
\begin{equation}
U(\omega )=\frac{\hbar \omega ^{3}}{\pi ^{2}c^{3}}\frac{1}{\exp (\hbar
\omega /k_{B}T)-1}
\end{equation}
yield the ratio of the spontaneous and stimulated emission coefficients
\begin{equation}
\frac{A}{B}=\frac{\hbar \omega ^{3}}{\pi ^{2}c^{3}}.
\end{equation}
When we deal with transitions in the XUV or X-ray regimes, the fast spontaneous decay rates, which are given by Einstein's $A$ coefficient, make it difficult to create population inversion. In the late 80's it was proposed~\cite{LWI} and demonstrated experimentally~\cite{LWIExp1,LWIExp2} that lasing can be achieved without population inversion if more than two levels are involved. This technique allows lasing even when a small fraction of population is in the excited state.

Our goal is to investigate the extent to which coherence effects can help to make shorter wavelength lasers in transient regime. Here we study how presence of a coherent drive at optical frequency, can enhance radiation generated in the adjacent XUV or X-ray lasing transition and, thus, utilize the advantages of the recombination excitation technique and the quantum coherence effects. We consider a three-level scheme and, as an example, will have in mind gas of He atoms or He-like Carbon ions as an active medium. The corresponding energy levels of He and C$^{4+}$ and their decay rates are shown in Fig. \ref{Fig5}. We assume that $a\leftrightarrow c$ optical transition is driven by a coherent resonant field with Rabi frequency $\Omega _{c}$ while the short wavelength transition $a\leftrightarrow b $ is coupled to a weak probe laser field $\Omega _{b}$\cite{J1}. We disregard contributions to decoherence caused by $T_{2}$ processes.
\section{Gain enhancement by coherent drive}
\begin{figure}[b]
\centerline{\includegraphics[height=5cm,width=0.30\textwidth,angle=0]{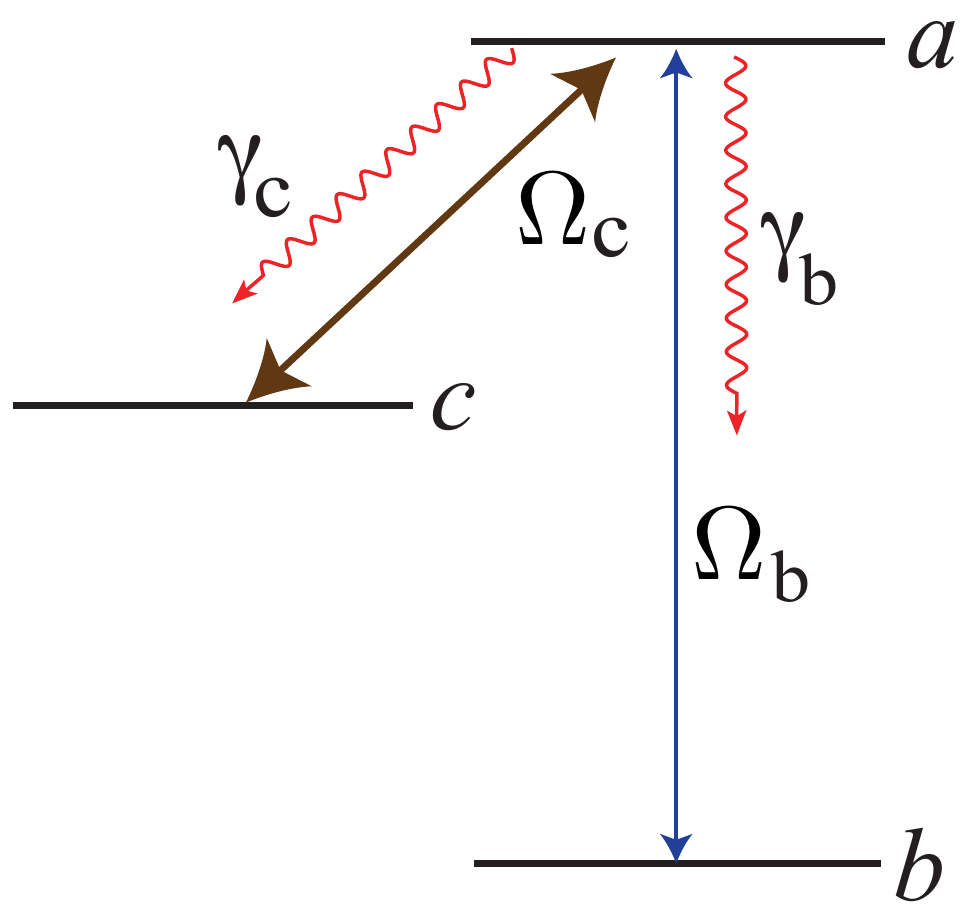}}
\caption{Three-level atomic system in $\Lambda -$configuration. }
\label{levels}
\end{figure}

We consider three-level atomic system in Lambda ($\Lambda )$ configuration where the transitions $a\leftrightarrow c$ and $a\leftrightarrow b$ are dipole allowed but the transition $c\leftrightarrow b$ is forbidden (see
Fig. \ref{levels}). We assume that at the initial moment of time the population is distributed between all three levels which can be achieved, e.g., by the ionization-recombination excitation. Transition $a\leftrightarrow c$ is driven in resonance with the Rabi frequency $\Omega_{c}$. We investigate how a weak laser seed pulse at the $a\leftrightarrow b$ transition evolves during its propagation through the medium. Evolution of
the atomic density matrix $\varrho _{ij}$ is described by the set of coupled equations~\cite{MOS}
\begin{equation}
\dot{\varrho}_{ab}=-\Gamma _{ab}\varrho _{ab}+i\Omega _{b}(\varrho_{bb}-\varrho _{aa})+i\Omega _{c}\varrho _{cb},  \label{eq3}
\end{equation}
\begin{equation}
\dot{\varrho}_{cb}=i(\Omega _{c}^{\ast }\varrho _{ab}-\Omega _{b}\varrho_{ac}^{\ast }),  \label{eq4}
\end{equation}
\begin{equation}
\dot{\varrho}_{ac}=-\Gamma _{ac}\varrho _{ac}-i\Omega _{c}(\varrho_{aa}-\varrho _{cc})+i\Omega _{b}\varrho _{cb}^{\ast },  \label{eq5}
\end{equation}
\begin{equation}
\dot{\varrho}_{aa}=-(\gamma _{c}+\gamma _{b})\varrho _{aa}-i\left( \Omega_{c}^{\ast }\varrho _{ac}-\text{c.c}\right) -i\left(\Omega _{b}^{\ast }\varrho _{ab}-\text{c.c}\right) ,
\label{eq6}
\end{equation}
\begin{equation}
\dot{\varrho}_{cc}=\gamma _{c}\varrho _{aa}+i(\Omega _{c}^{\ast }\varrho_{ac}-\text{c.c}),  \label{eq7}
\end{equation}
\begin{equation}
\varrho _{aa}+\varrho _{bb}+\varrho _{cc}=1,  \label{eq8}
\end{equation}
where $\Gamma _{ab}=\Gamma _{ac}=(\gamma _{c}+\gamma _{b})/2$ are the relaxation rates of the off-diagonal elements of the atomic density matrix, $\gamma _{c}$ and $\gamma _{b}$ are the spontaneous decay rates into the levels $c$ and $b$, $\Omega _{c}$ is the Rabi frequency of the laser field.

Next we discuss two regimes, namely $\gamma _{c}\gg \gamma _{b}$ which can be treated analytically and $\gamma _{c}\lesssim \gamma _{b}$ which we investigate numerically.
\subsection{Steady-state approximation: $\protect\gamma_{c} \gg \protect\gamma_{b}$}
Here we assume that $\gamma _{c}\gg \gamma _{b}$ and populations in the levels $a$ and $c$ reach approximate steady state (that is $1/\gamma _{b}\gg t\gg 1/\gamma _{c}$). To find analytical solution we assume that $\Omega _{c}=$const (a real number) and $\Omega_{b}$ is very small. Under these assumptions equations describing evolution of levels $c$ and $a$ decouple and become
\begin{equation}
\dot{\varrho}_{cc}=\gamma _{c}\varrho _{aa}+i\Omega _{c}(\varrho_{ac}-\text{c.c}),  \label{g11}
\end{equation}
\begin{equation}
\dot{\varrho}_{aa}=-(\gamma _{c}+\gamma _{b})\varrho _{aa}-i\Omega_{c}(\varrho _{ac}-\text{c.c}),  \label{g12}
\end{equation}
\begin{equation}
\dot{\varrho}_{ac}=-\Gamma_{ac}\varrho_{ac}-i\Omega _{c}(\varrho _{aa}-\varrho _{cc}).  \label{g13}
\end{equation}
The steady state solution ($\bar{\varrho}_{ij} $) of these equations is (we put $\gamma _{b}=0$)
\begin{equation}
\bar{\varrho} _{aa}=\frac{4\Omega _{c}^{2}}{\gamma _{c}^{2}+8\Omega _{c}^{2}}[\varrho _{cc}(0)+\varrho _{aa}(0)],  \label{g57}
\end{equation}
\begin{equation}
\bar{\varrho}  _{cc}=\frac{\gamma _{c}^{2}+4\Omega _{c}^{2}}{\gamma_{c}^{2}+8\Omega _{c}^{2}}[\varrho _{cc}(0)+\varrho _{aa}(0)],  \label{g58}
\end{equation}
\begin{equation}
\bar{\varrho} _{ac}=\frac{2i\gamma _{c}\Omega _{c}}{\gamma _{c}^{2}+8\Omega_{c}^{2}}[\varrho _{cc}(0)+\varrho _{aa}(0)],  \label{g59}
\end{equation}
where $\varrho _{cc}(0)+\varrho _{aa}(0)$ is the net population of the levels $c$ and $a$.  Evolution of the weak laser pulse $\Omega _{b}$ is described by the Maxwell equations which in the slowly varying amplitude approximation can be written as
\begin{equation}
\frac{\partial \Omega _{b}}{\partial z}+\frac{1}{c}\frac{\partial \Omega _{b}}{\partial t}=i\eta_{ab} \varrho _{ab},  \label{g60}
\end{equation}
where $\eta_{ab}  =(3/8\pi )N\lambda _{ab}^{2}\gamma _{b}$ is the coupling constant, $N$ is the atomic density and $\lambda _{ab}$ is the wavelength of the $a\leftrightarrow b$ transition. This equation must be supplemented by the equation for $\varrho _{ab}$
\begin{equation}
\dot{\varrho}_{ab}=-\Gamma_{ab}\varrho _{ab}+i\Omega_{b}(\bar{\varrho} _{bb}-\bar{\varrho}_{aa})+i\Omega _{c}\varrho _{cb}  \label{g5}
\end{equation}
which couples to the equation for $\rho _{cb}$
\begin{equation}
\dot{\varrho}_{cb}=i(\Omega _{c}\varrho _{ab}-\Omega _{b}\bar{\varrho} _{ac}^{\ast }).
\label{g6}
\end{equation}
Here we took into account that $\varrho _{bb},$ $\varrho _{aa}$ and $\varrho _{ac}$ are approximately constant. Let us look for solution of Eqs. (\ref{g60}-\ref{g6}) in the form of a plain wave
\begin{equation}
\Omega _{b}(t,z)\sim e^{i\omega t-ikz}  \label{g63}
\end{equation}
\begin{equation}
\varrho _{ab}(t,z)\sim e^{i\omega t-ikz}  \label{g63a}
\end{equation}
\begin{equation}
\varrho_{cb}(t,z)\sim e^{i\omega t-ikz}  \label{g63b}
\end{equation}
which yields the following dispersion relation
\begin{equation}
\begin{split}
\left( \omega ^{2}-\Omega _{c}^{2}-\frac{i\gamma _{c}\omega }{2}\right)&\left( ck-\omega \right) +c\omega \eta _{ab}(\bar{\varrho}  _{bb}-\bar{\varrho} _{aa})\\
&+c\eta _{ab}\Omega _{c}\bar{\varrho} _{ac}=0,
\end{split}
\label{g64}
\end{equation}
here $\omega $ is the detuning of the laser pulse frequency from the $a\leftrightarrow b$ transition frequency. If in Eq. (\ref{g64}) we treat $\omega $ as real then imaginary part of $k$ gives gain (absorption) per unit length as a function
of $\omega $
\begin{equation}
\text{Im}(k)=\eta _{ab}\frac{\gamma _{c}\omega ^{2}(\bar{\varrho}  _{aa}-\bar{\varrho} _{bb})/2+\Omega _{c}\left( \Omega _{c}^{2}-\omega ^{2}\right) \text{Im}(\bar{\varrho}_{ac})}{\left( \omega ^{2}-\Omega _{c}^{2}\right) ^{2}+\gamma_{c}^{2}\omega ^{2}/4}.
\label{m1}
\end{equation}
In particular, for the mode resonant with the $a\leftrightarrow b$ transition $\omega =0$ and we obtain
\begin{equation}
G=\text{Im}(k)=\frac{\eta _{ab}}{\Omega _{c}}\text{Im}(\bar{\varrho}_{ac}).
\label{m2}
\end{equation}
Eq. (\ref{m2}) shows that if Im$(\bar{\varrho} _{ac})>0$ there is positive gain no matter what are the populations of the levels $a$ and $b$. Thus, one can have gain without population inversion. This is the case for the $\Lambda -$ scheme in the approximate steady state for which, according to Eq. (\ref{g59}), Im$(\bar{\varrho} _{ac})>0$. However, in the transient regime the steady state approximation is valid only for $\gamma _{c}\gg \gamma _{b}$. If $\gamma _{c}\lesssim \gamma _{b}$ the time evolution of atomic populations must be taken into account. In this regime we found no gain without population inversion. However, presence of the coherent drive field $\Omega _{c}$ can enhance lasing with inversion. We discuss this next.
\subsection{Coherence enhanced transient lasing with population inversion}
\begin{figure}[t]
\centerline{\includegraphics[height=4.4cm,width=0.49\textwidth,angle=0]{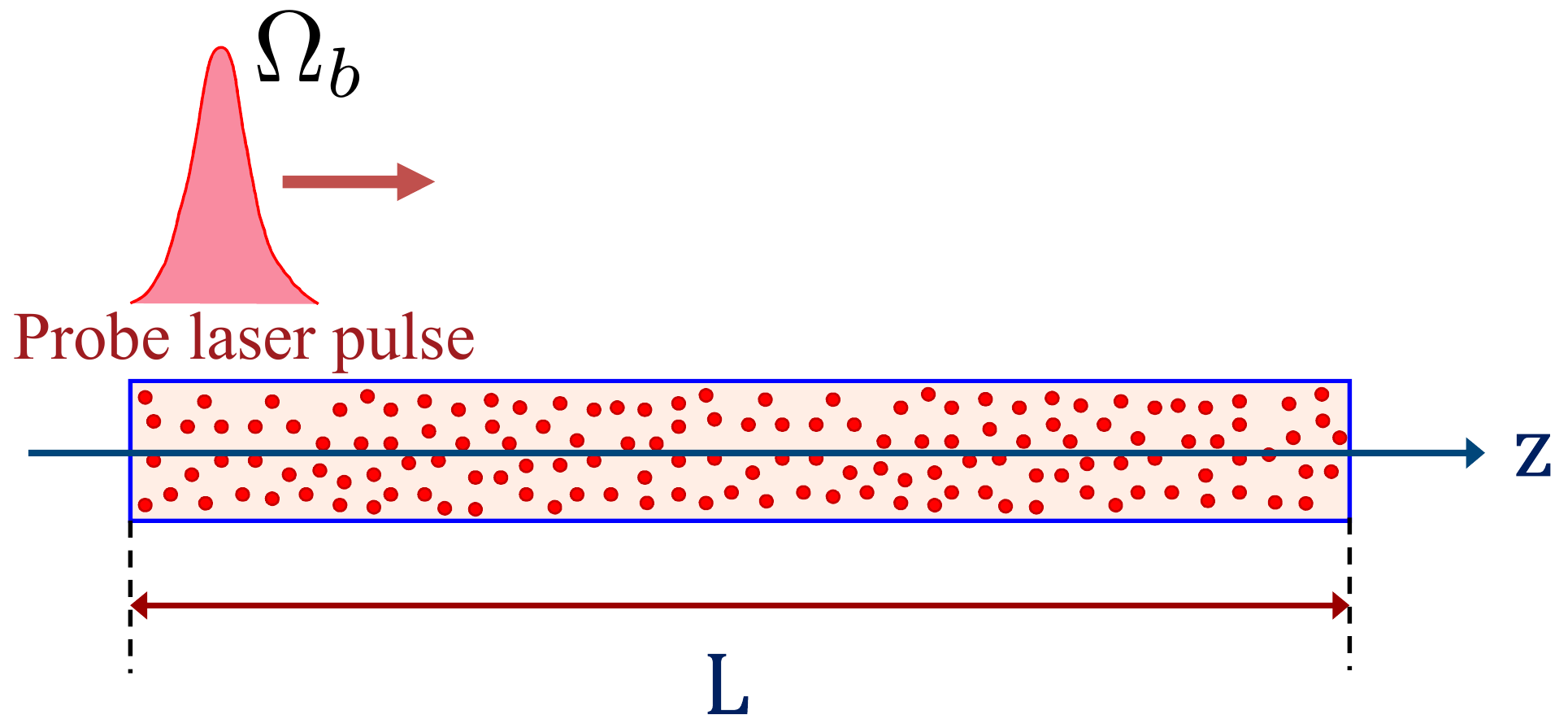}}
\caption{Weak laser probe pulse $\Omega_b$ propagates through the atomic medium of length $L$ gaining or losing its energy.}
\label{Fig6}
\end{figure}

As before, we consider a three-level scheme having in mind gas of He atoms or He-like Carbon ions as an active medium. The corresponding energy levels and their decay rates are shown in Fig. \ref{Fig5}. We are interested in evolution of a weak laser pulse $\Omega _{b}(t,z)$ propagating along the $z-$axis through the atomic medium (see Fig. \ref{Fig6}). First we discuss the $\Lambda $-scheme shown in Fig. \ref{levels}. We assume that driving field $\Omega _{c}=$const, however, populations of the levels $a$, $b $ and $c$ depend on time (transient regime). We use semiclassical approach in which evolution of $\Omega _{b}(t,z)$ is described by the Maxwell's equation (\ref{g60}) which is supplemented by the quantum mechanical equations (\ref{eq3})-(\ref{eq8}) for the atomic density matrix.
\begin{figure}[b]
\centerline{\includegraphics[height=5.6cm,width=0.47\textwidth,angle=0]{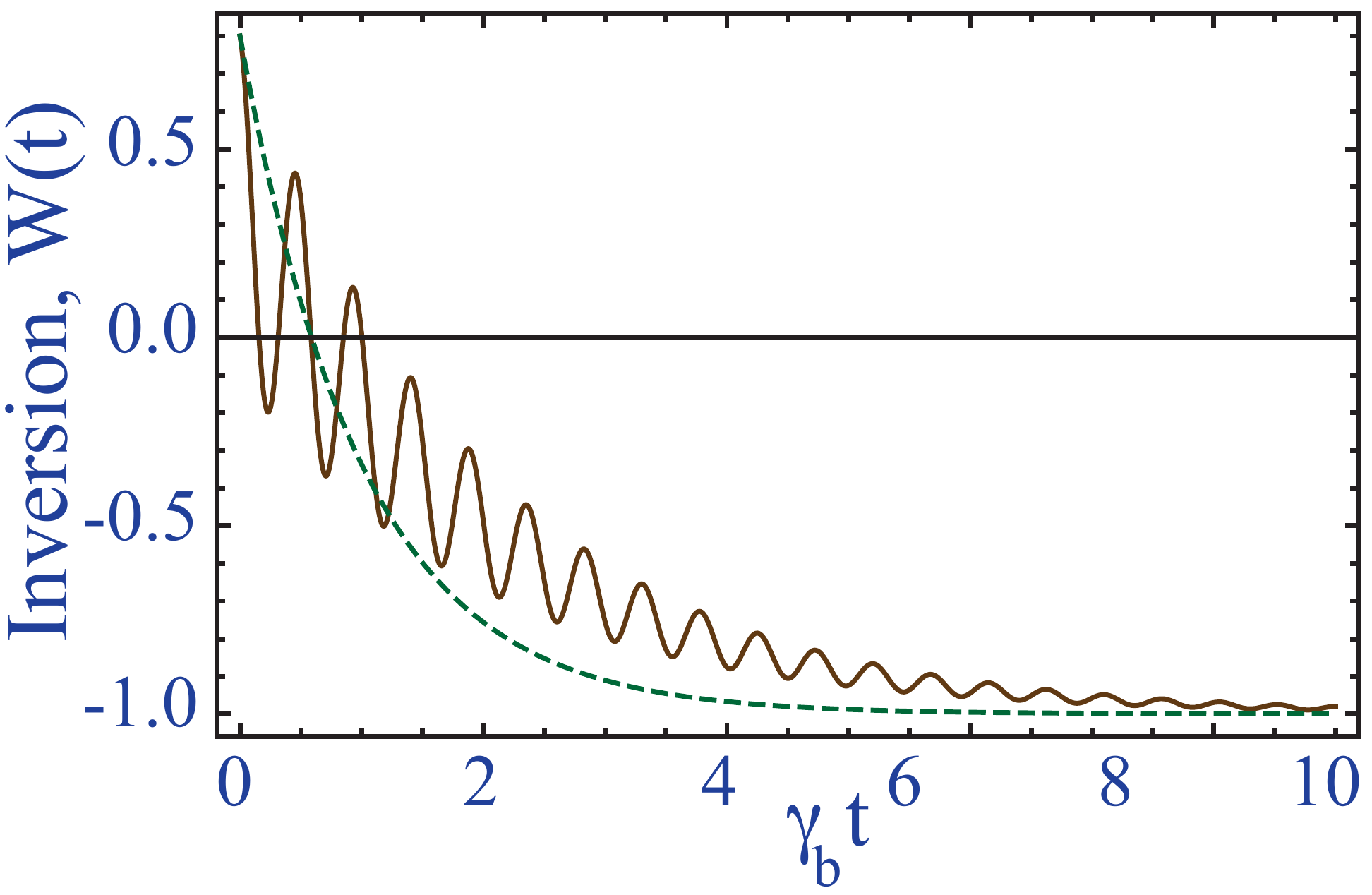}}
\caption{Inversion $W(t)$ in the probe transition ($a\leftrightarrow b$) vs dimensionless time $\protect\gamma _{b}t$. Solid curve shows the result for $\Omega _{c}=6.625\protect\gamma _{b}$ while dashed line is obtained with no drive. In calculations we take $\protect\gamma _{c}=1.83\times 10^{-5}\protect\gamma _{b}$ and the initial condition $\protect\varrho _{aa}(0)=0.9,$ $\protect\varrho _{bb}(0)=0.1$, $\protect\varrho _{cc}(0)=\protect\varrho_{ac}(0)=\protect\varrho _{ab}(0)=0$}
\label{Fig7}
\end{figure}
For a weak probe field $\Omega _{b}$ one can put $\Omega _{b}=0$ in Eqs. (\ref{eq5})-(\ref{eq7}). Then Eqs. (\ref{eq5})-(\ref{eq7}) for the density matrix elements $\varrho _{ac},$ $\varrho _{aa}$ and $\varrho _{cc}$ decouple from the other equations. In particular, for $\Omega _{c},\gamma_{b}\gg \gamma _{c}$ and $\varrho _{ac}(0)=0$ we obtain (assuming $\Omega_{c}$ is real)
\begin{equation}  \label{eq9}
\begin{split}
\varrho _{aa} =e^{-\gamma _{b}t/2}\varrho _{aa}(0)\left\{ \left(1+\frac{\varrho _{cc}(0)}{\varrho _{aa}(0)}\right)\sin^{2}(\Omega _{c}t)+\right.\\
\left. \cos (2\Omega _{c}t)-\frac{\gamma _{b}}{4\Omega _{c}}\sin (2\Omega_{c}t)\right\},
\end{split}
\end{equation}
\begin{equation}  \label{eq10}
\begin{split}
\varrho _{cc} =e^{-\gamma _{b}t/2}\varrho _{aa}(0)\left\{ \left(1+\frac{\varrho _{cc}(0)}{\varrho _{aa}(0)}\right)\sin^{2}(\Omega _{c}t)+\right.\\
\left. \frac{\varrho _{cc}(0)}{\varrho _{aa}(0)}\cos (2\Omega _{c}t)+\frac{\gamma _{b}}{4\Omega _{c}}\frac{\varrho _{cc}(0)}{\varrho _{aa}(0)}\sin(2\Omega _{c}t)\right\},
\end{split}
\end{equation}
\begin{figure}[b]
\centerline{\includegraphics[height=5cm,width=0.47\textwidth,angle=0]{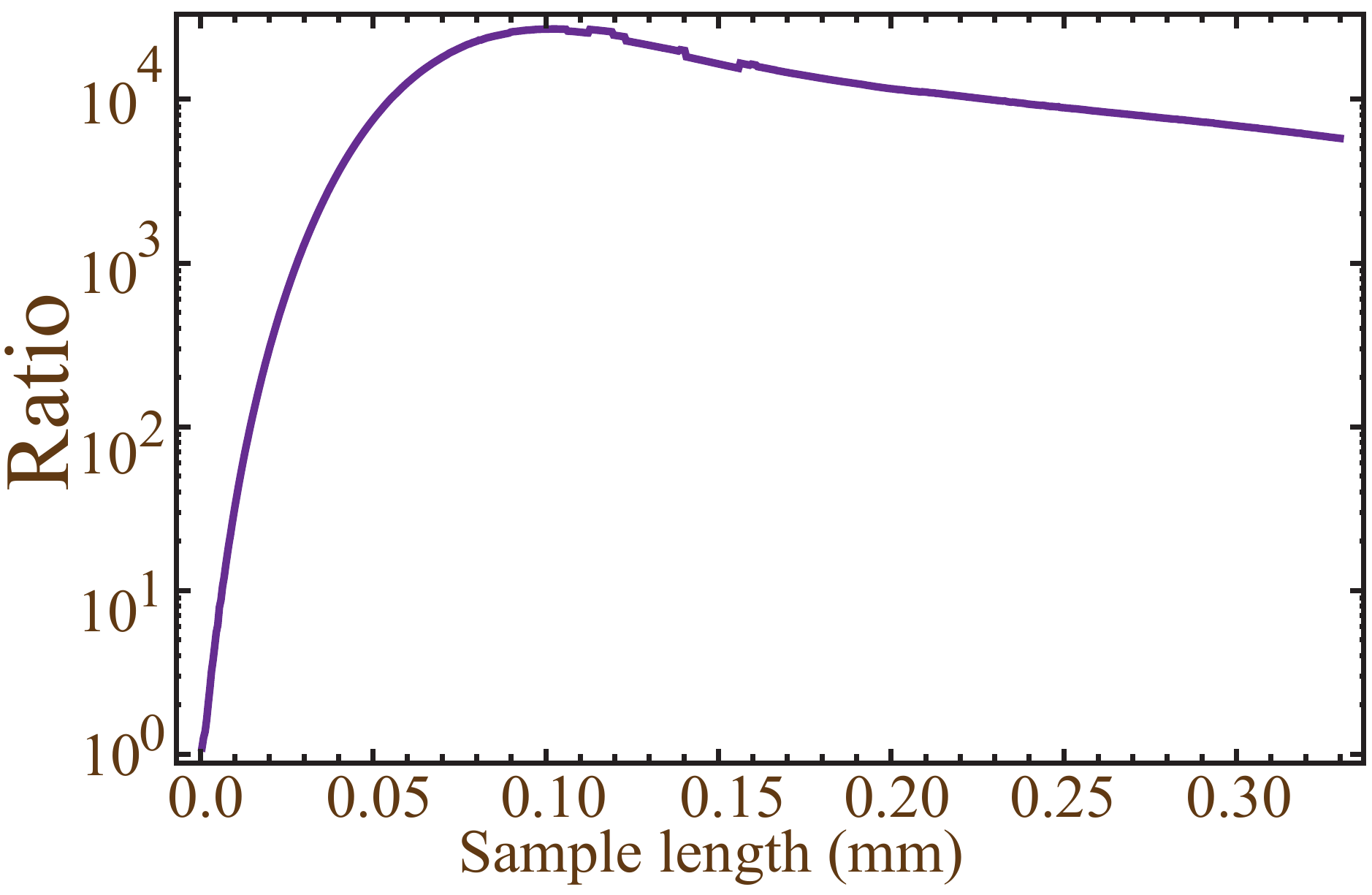}}
\caption{Ratio of the output energy to the input energy of the probe laser pulse as a function of sample length $L$ with no external drive. In numerical simulations we take $\protect\gamma _{c}=1.83\times 10^{-5}\protect%
\gamma _{b}$, $\protect\eta /\protect\gamma _{b}=19353$ cm$^{-1}$ and assume Gaussian initial probe pulse shape given by Eq. (\protect\ref{eq13}). Initial populations are $\protect\varrho _{aa}(0)=0.9,$ $\protect\varrho %
_{bb}(0)=0.1$ and $\protect\varrho _{cc}(0)=0 $, while initial coherences are equal to zero.}
\label{Fig8}
\end{figure}
\begin{equation}  \label{eq11}
\begin{split}
\varrho _{ac}=ie^{-\gamma _{b}t/2}\varrho _{aa}(0)\sin (\Omega _{c}t)\left\{\left(\frac{\varrho _{cc}(0)}{\varrho _{aa}(0)}-1\right)\cos (\Omega_{c}t)\right. \\
\left. +\frac{\gamma _{b}}{4\Omega _{c}} \left(1+\frac{\varrho_{cc}(0)}{\varrho _{aa}(0)}\right)\sin(\Omega _{c}t)\right\}.
\end{split}
\end{equation}
Using Eq. (\ref{eq8}) for conservation of the net population we find that population difference between levels $a$ and $b$, defined as $W(t)=\varrho_{aa}(t)-\varrho _{bb}(t)$, is given by
\begin{equation}  \label{eq12}
\begin{split}
W(t)&=\frac{\varrho _{aa}(0)}{2}e^{-\gamma _{b}t/2}\left[ 3\left(1+\frac{\varrho _{cc}(0)}{\varrho _{aa}(0)}\right)+\left( 1-\frac{\varrho _{cc}(0)}{\varrho _{aa}(0)}\right)\right. \\
&\times \cos(2\Omega _{c}t)-\left. \frac{\gamma _{b}}{2\Omega _{c}}\left( 2-\frac{\varrho _{cc}(0)}{\varrho_{aa}(0)}\right) \sin (2\Omega _{c}t)\right]-1.
\end{split}
\end{equation}
\begin{figure}[t]
\centerline{\includegraphics[height=12cm,width=0.46\textwidth,angle=0]{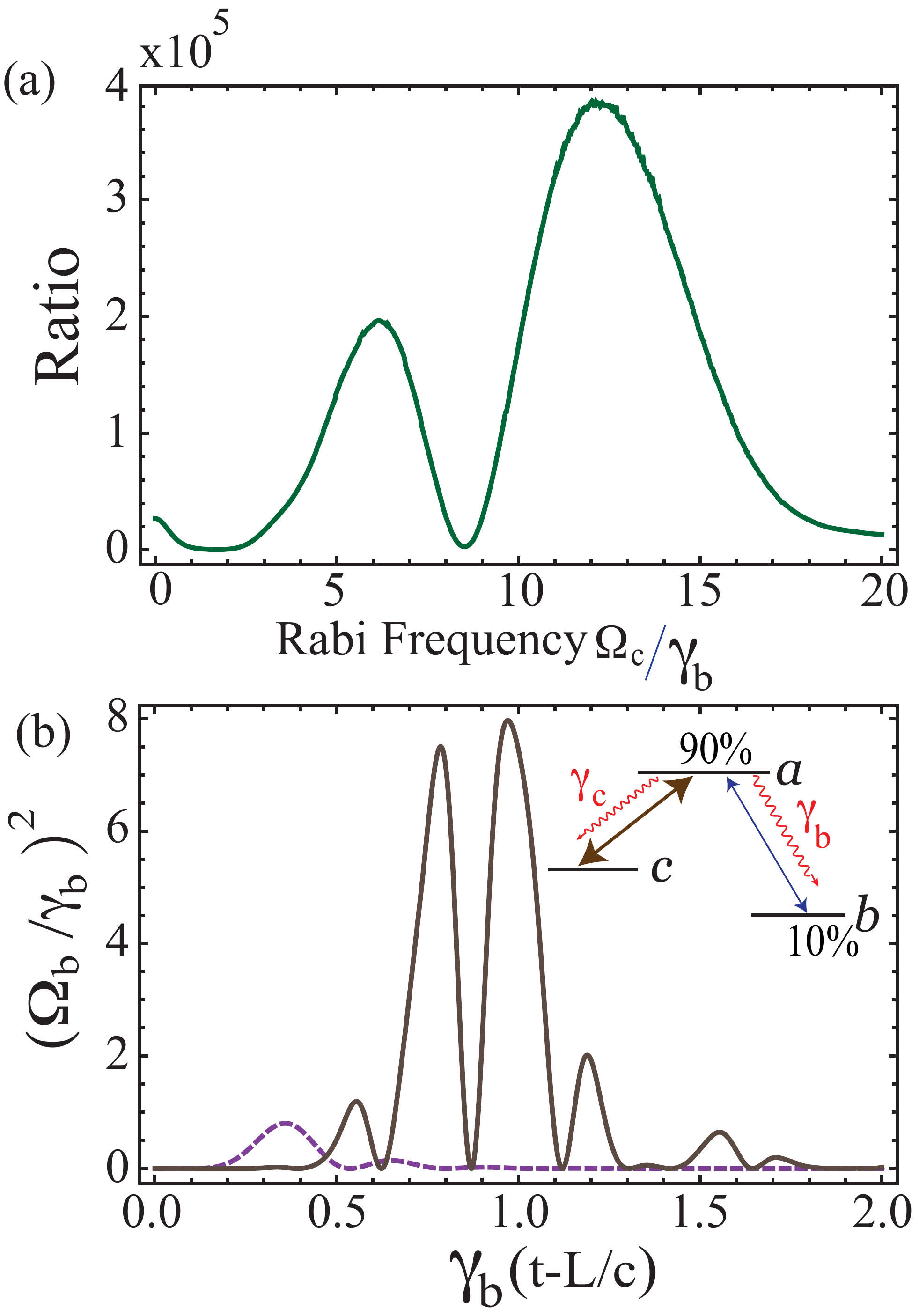}}
\caption{(a) Ratio of the output energy to the input energy of the probe laser pulse as a function of the driving field Rabi frequency $\Omega _{c}$.
The ratio is $\sim 2.7 \times 10^{4}$ at $\Omega_{c}=0$. (b) Square of the output probe pulse $\Omega _{b}/\protect\gamma _{b}$ as a function of time for optimal sample length $L=0.102$ mm with (solid line) and without
(dashed) coherent drive field $\Omega_{c}$. In numerical simulations we take $\protect\gamma _{c}=1.83\times 10^{-5}\protect\gamma _{b}$, $\protect\eta /\protect\gamma _{b}=19353$ cm$^{-1}$ and assume Gaussian initial probe pulse shape given by Eq. (\protect\ref{eq13}). The length of the sample is $L=0.102$ mm, while the initial populations are $\protect\varrho _{aa}(0)=0.9,$ $\protect\varrho _{bb}(0)=0.1 $, $\protect\varrho_{cc}(0)=0$ and $\protect\varrho _{ac}(0)=\protect\varrho _{ab}(0)=0$. }
\label{Fig9}
\end{figure}
In Fig. \ref{Fig7} we plot the population difference $W(t)$ as a function of time for initial conditions $\varrho _{aa}(0)=0.9,$ $\varrho _{bb}(0)=0.1$ and $\varrho_{cc}(0)=0$. Solid line is obtained for $\Omega _{c}=6.625\gamma_{b}$ while for dashed line $\Omega _{c}=0$. Driving the $a\leftrightarrow c$ transition yields oscillations in the population difference between $a$ and $b$ levels.
\subsubsection{Helium-like Carbon}
Next we solve Eqs. (\ref{eq3})-(\ref{eq8}) and (\ref{g60}) numerically and obtain evolution of the probe laser pulse $\Omega _{b}(t,z)$ when the $a\leftrightarrow c$ transition is driven by a constant coherent field $\Omega _{c}$ or by a constant incoherent pump $\Phi $. We perform simulations for the initial condition $\varrho _{aa}(0)=0.9,$ $\varrho_{bb}(0)=0.1,$ $\varrho _{cc}(0)=0$ and take $\eta /\gamma _{b}=19353$ cm$^{-1}$ and $\gamma _{c}=1.83\times 10^{-5}\gamma _{b}$. As an example, we consider He-like Carbon ions for which states 2$^{1}$S$_{0}$ ($c-$ level), 2$^{1}$P$_{1}$ ($a-$level) and the ground state 1$^{1}$S$_{0}$ ($b-$level) form $\Lambda -$scheme (see Fig. \ref{Fig5}). For C$^{4+}$ ions the model parameters are $\lambda _{ab}=4.027$ nm, $\lambda _{ac}=352.5$ nm, $\gamma_{c}=1.67\times 10^{7}$ s$^{-1}$ and $\gamma _{b}=9.09\times 10^{11}$ s$^{-1}$. Then for ion density $N=10^{18}$ cm$^{-3}$ we obtain $\eta /\gamma_{b}=19353$ cm$^{-1}$.
We assume that input probe laser pulse has a Gaussian shape
\begin{equation}  \label{eq13}
\Omega _{b}(t,z=0)=0.01\exp \left[ -\left( \frac{\gamma _{b}t-0.15}{0.05}\right) ^{2}\right] \gamma _{b}.
\end{equation}
During propagation of the weak laser pulse through the medium the atomic population spontaneously decays into the ground state. After a certain time the medium is no longer inverted and the laser pulse begins to attenuate. Thus, there is an optimum length of the atomic sample which yields maximum enhancement of the pulse energy. For the optimum length the pulse leaves the medium at the onset of absorption. In Fig. \ref{Fig8} we plot the ratio of the output pulse energy to the input energy as a function of the sample length assuming there is no external drive. We find that optimum length corresponding to maximum output energy without any drive is $L=0.102$ mm. At
this optimum length the ratio of the output to the input probe field energy is $\sim 2.7\times 10^{4}$.

Next we turn on the coherent driving field $\Omega _{c}$, but keep the sample length to be $L=0.102$ mm. This length does not corresponds to the maximum gain for three-level system and chosen as a demonstration that
coherent drive can enhance the gain for a fixed sample size. In Fig. \ref{Fig9}$a$ we plot the ratio of the output laser pulse energy (at $z=L$) to the input energy (at $z=0$) as a function of strength of the driving field $\Omega _{c}$. One can see that in the presence of coherent drive the output pulse energy oscillates as a function of $\Omega _{c}$. Such oscillations appear because coherence $\varrho _{ac}$ averaged over the pulse propagation time depends on $\Omega _{c}$. At $\Omega _{c}\sim 6\gamma _{b}$ the enhancement factor is $7$ as compared to the case with no drive field. The enhancement factor increases upto $14$ for $\Omega _{c}\sim 12\gamma _{b}$. Thus, coherent drive can increase the laser pulse output energy more than an order of magnitude as compared to the pulse energy with no drive. Fig. \ref{Fig9}b shows the shape of the output pulse $\Omega _{b}(t,z=L)$ for the optimum length in the absence of the external drive (dashed line) and optimum coherent drive $\Omega _{c}=12\gamma _{b}$ (solid line).
\begin{figure}[t]
\centerline{\includegraphics[height=5.5cm,width=0.46\textwidth,angle=0]{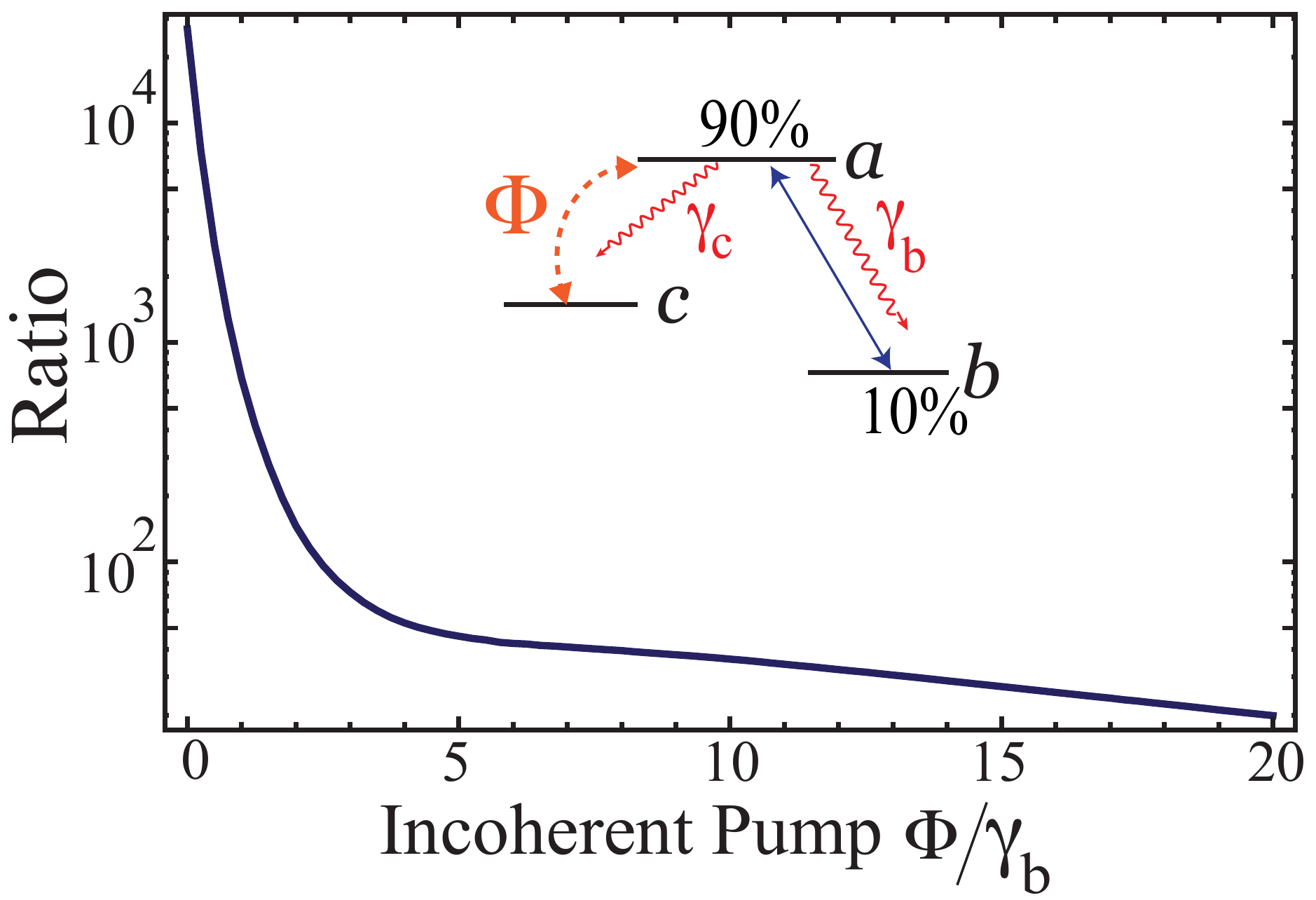}}
\caption{Ratio of the output energy to the input energy of the probe laser pulse as a function of the incoherent pump rate $\Phi$. The ratio is $\sim2.7 \times 10^{4}$ at $\Phi=0$}
\label{Fig10}
\end{figure}

If we replace the coherent drive by an incoherent pump $\Phi $, which does not induce coherence, the gain becomes smaller when $\Phi $ increases (see
Fig. \ref{Fig10}). The enhancement due to coherence can also be obtained for He gas as an active medium at much lower density $N=10^{13}$ cm$^{-3}$ with lasing at $58.4$ nm. Now the model parameters are given in the left side of Fig. \ref{Fig5} which yields $\eta /\gamma _{b}=40.75$ cm$^{-1}$.
\subsubsection{Neutral He as active medium}
\begin{figure}[t]
\centerline{\includegraphics[height=11cm,width=0.46\textwidth,angle=0]{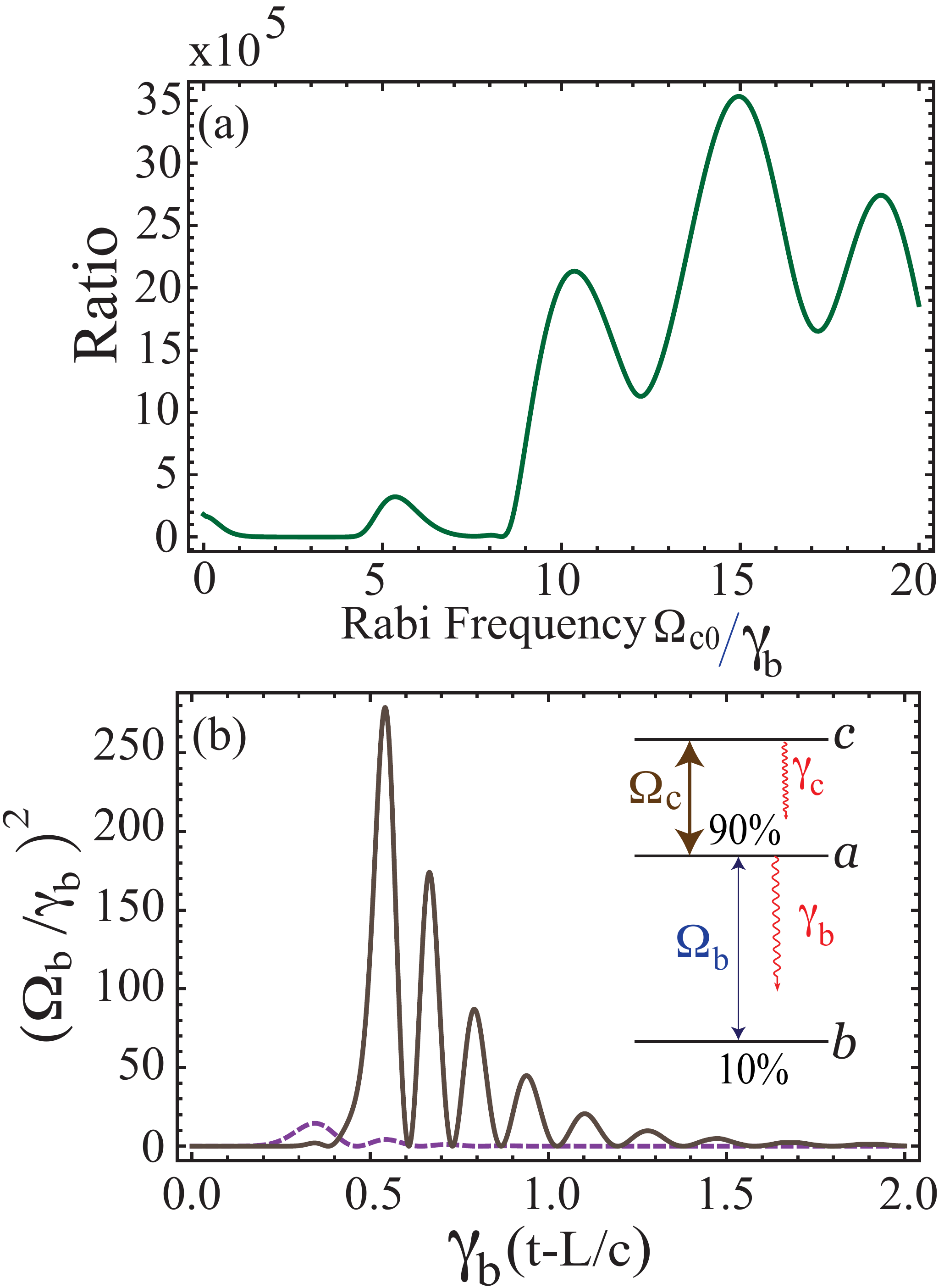}}
\caption{(a) Ratio of the output energy to the input energy of the probe laser pulse as a function of the driving field Rabi frequency $\Omega _{c0}$. (c) Square of the output probe pulse $\Omega _{b}/\protect\gamma _{b}$ as
a function of time with $\Omega_{c0}=15 \protect\gamma_b$ (solid line) and $\Omega_c=0$ (dashed line). In numerical simulations we take $\protect\gamma _{c}=0.01\protect\gamma _{b}$, $\protect\eta /\protect\gamma _{b}=81.50$ cm$^{-1}$ and assume Gaussian initial probe laser pulse (Eq. (\protect\ref{eq14})) and Gaussian driving field (Eq. (\protect\ref{eq15})). The length of the sample is $L=5.19$ cm, while the initial populations are $\protect\varrho_{aa}(0)=0.9,$ $\protect\varrho _{bb}(0)=0.1$, $\protect\varrho _{cc}(0)=0$ and $\protect\varrho _{ca}(0)=\protect\varrho _{ab}(0)=0$. }
\label{Fig11}
\end{figure}
Next we consider $\Xi -$scheme formed by the 3$^{1}$S$_{0}$ ($c-$ level), 2$^{1}$P$_{1}$ ($a-$level) and the ground state 1$^{1}$S$_{0}$ ($b-$level) of the Helium atom (see Fig. \ref{Fig5} right side). For this scheme the model parameters are $\lambda _{ab}=58.4$ nm, $\lambda _{ca}=728.3$ nm, $\gamma_{c}=1.83\times 10^{7}$ s$^{-1}$ and $\gamma _{b}=1.82\times 10^{9}$ s$^{-1}$. Then for atomic density $N=2\times 10^{13}$ cm$^{-3}$ we obtain $\eta/\gamma _{b}=81.50$ cm$^{-1}$. We assume that the input probe laser pulse has a Gaussian shape
\begin{equation}
\Omega _{b}(t,z=0)=0.01\exp \left[ -\left( \frac{\gamma _{b}t-0.28}{0.10}\right) ^{2}\right] \gamma _{b},  \label{eq14}
\end{equation}
while the drive pulse is also Gaussian with a broader width
\begin{equation}  \label{eq15}
\Omega _{c}(t,z=0)=\Omega_{c0}\exp \left[ -\left( \frac{\gamma _{b}t-0.28}{0.40}\right) ^{2}\right].
\end{equation}
Similar to the $\Lambda $-scheme we first optimize the length of the sample for the given initial population distribution and obtain that the optimum length corresponding to maximum output energy without drive is $L=5.19$ cm. Then we turn on the driving field $\Omega _{c}$, but keep the sample length to be the same. Fig. \ref{Fig11}$a$ shows the ratio of the output laser pulse energy (at $z=L$) to the input energy (at $z=0$) as a function of strength of the coherent drive. One can see that, similar to the $\Lambda $ configuration, the output pulse energy oscillates as a function of $\Omega_{c}$ and the laser pulse output energy can be increased more than an order of magnitude as compared to the pulse energy with no drive. Thus, coherence can help to extract more energy from the inverted medium and convert it into coherent laser radiation for both $\Lambda $ and cascade configurations. Fig. \ref{Fig11}$b$ shows the shape of the output pulse $\Omega _{b}(t,z=L)$ for $\Omega _{c0}=0$ (dashed line) and optimum coherent drive of $\Omega_{c0}=15\gamma _{b}$ (solid line).

\section{Backward vs Forward gain }
Let us now consider the same problem $(\Lambda$-scheme), with focus on controlling the gain in the forward (with respect to the pump pulse) and the backward direction. We will use the superscript $(+)$ for forward and $(-)$ for backward direction. The density matrix equations for the populations and the coherence can be written as \cite{Comment1,Comment2}
\begin{equation}
\dot{\varrho}_{ab}^{+}=-\Gamma_{ab}\varrho _{ab}^{+}+i\Omega_{b}^{+}(\varrho _{bb}-\varrho _{aa})+i\Omega _{c}\varrho _{cb}^{+},  \label{CETLeq35}
\end{equation}
\begin{equation}
\dot{\varrho}_{ab}^{-}=-\Gamma_{ab}\varrho _{ab}^{-}+i\Omega_{b}^{-}(\varrho _{bb}-\varrho _{aa}), \label{CETLeq36}
\end{equation}
\begin{equation}
\dot{\varrho}_{ac}^{+}=-\Gamma_{ac}\varrho _{ac}^{+}-i\Omega_{c}(\varrho _{aa}-\varrho _{cc})+i\Omega _{b}^{+}\varrho _{cb}^{+\ast },  \label{CETLeq37}
\end{equation}
\begin{equation}
\dot{\varrho}_{ac}^{-}=-\Gamma_{ac}\varrho _{ac}^{-}+i\Omega _{b}^{-}\varrho _{cb}^{-\ast },  \label{CETLeq38}
\end{equation}
\begin{equation}
\dot{\varrho}_{cb}^{+}=i(\Omega _{c}^{+\ast }\varrho _{ab}^{+}-\Omega _{b}^{+}\varrho _{ac}^{+\ast}),  \label{CETLeq39}
\end{equation}
\begin{equation}
\dot{\varrho}_{cb}^{-}=-\Omega _{b}^{-}\varrho _{ac}^{-\ast},  \label{CETLeq40}
\end{equation}
\begin{equation}
\begin{split}
\dot{\varrho}_{aa}=-(\gamma _{c}+\gamma _{b})\varrho _{aa}-i\left( \Omega_{c}^{+\ast }\varrho _{ac}^{+} +\Omega_{b}^{+\ast }\varrho _{ab}^{+}\right. \\
\left.+\Omega _{b}^{-\ast }\varrho _{ab}^{-}-\text{c.c.}\right)
\end{split}
\label{CETLeq41}
\end{equation}
\begin{equation}
\dot{\varrho}_{cc}=\gamma _{c}\varrho _{aa}+i(\Omega _{c}^{+\ast }\varrho _{ac}^{+}+\text{c.c}),  \label{CETLeq42}
\end{equation}
\begin{equation}
\varrho _{aa}+\varrho _{bb}+\varrho _{cc}=1,  \label{CETLeq43}
\end{equation}
where $\Gamma_{ab}=\Gamma_{ac}=(\gamma_{b}+\gamma_{c})/2$. The evolution of the backward and forward laser pulse $\Omega_{b}^{\pm}$ is governed by
\begin{equation}\label{CETLeq44}
\frac{\partial \Omega _{b}^{+}}{\partial z}+\frac{1}{c}\frac{\partial \Omega _{b}^{+}}{\partial t}=i\eta_{ab}\varrho _{ab}^{+}, 
\end{equation}
\begin{equation}\label{CETLeq45}
-\frac{\partial \Omega _{b}^{-}}{\partial z}+\frac{1}{c}\frac{\partial \Omega _{b}^{-}}{\partial t}=i\eta_{ab}\varrho _{ab}^{-}.
\end{equation}
For a two level system $(\Omega_{c}=0)$, the forward and backward directions are symmetric, i.e  transformation $+\leftrightarrow -$ yields the same equations. Thus $\Omega^{+}_{b}$ and $\Omega^{-}_{b}$ will be identical. This symmetry is broken in the presence of a coherent drive.

\begin{figure}[t]
\centerline{\includegraphics[height=5.5cm,width=0.47\textwidth,angle=0]{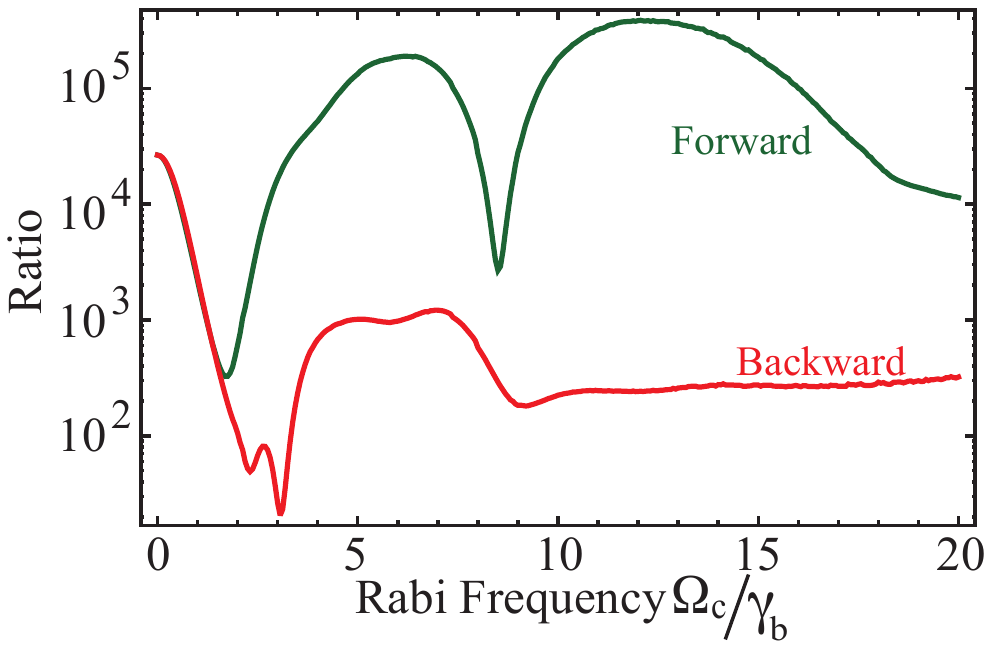}}
\caption{Ratio of the output energy to the input energy of the probe laser pulse as a function of the driving field Rabi frequency $\Omega _{c}$. The ratio is $\sim 2.65 \times 10^{4}$ at $\Omega_{c}=0$.  In numerical simulations we take $\protect\gamma _{c}=1.83\times 10^{-5}\protect\gamma _{b}$, $\protect\eta /\protect\gamma _{b}=19353$ cm$^{-1}$ and sample length $L=102.21\,\mu$m. We assumed Gaussian initial probe pulse shape given by Eq. (\protect\ref{eq13}) for both forward and backward direction. The initial populations are $\protect\varrho _{aa}(0)=0.9,$ $\protect\varrho _{bb}(0)=0.1$,  $\protect\varrho_{cc}(0)=0$ and the initial coherence in both forward and backward direction $\protect\varrho _{ac}(0)=\protect\varrho _{ab}(0)=\protect\varrho _{cb}(0)$.}
\label{CETLFig10}
\end{figure}
\begin{figure}[t]
\centerline{\includegraphics[height=5.0cm,width=0.48\textwidth,angle=0]{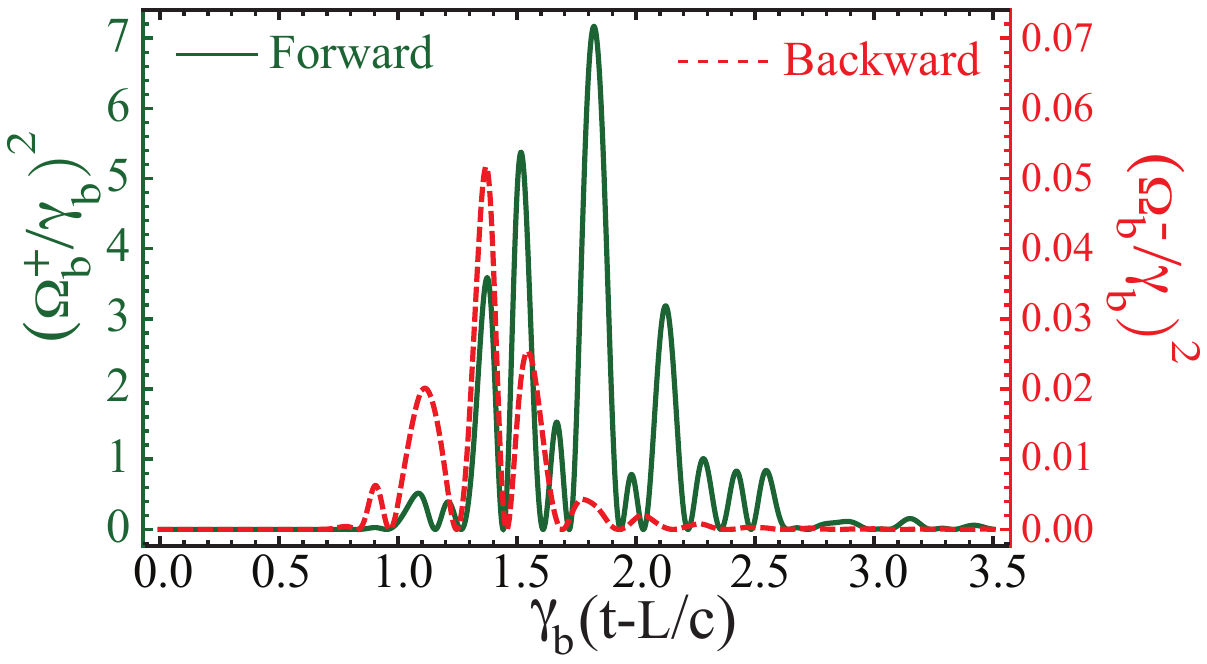}}
\caption{Square of the output probe pulse $\Omega _{b}/\protect\gamma _{b}$ as a function of time $\gamma_{b}t$. In numerical simulations we take the same parameters as in Fig. 10 with $\Omega_{c}=12 \gamma_{b}$.}
\label{CETLFig11}
\end{figure}

Next we solve Eqs. (\ref{CETLeq35})-(\ref{CETLeq45}) numerically and obtain evolution of the probe laser pulse $\Omega _{b}(t,z)$ in forward and backward directions when $a\leftrightarrow c$ transition is driven by a constant coherent field $\Omega _{c}$. We assume the input probe laser pulse has Gaussian shape given by Eq.(\ref{eq13}). In Fig. \ref{CETLFig10} we plot the ratio of the output energy to input energy of the probe pulse in the forward and backward directions as a function of the coherent drive field $\Omega_{c}$. Contrary to the forward direction, for which forward gain can be enhanced for some choice of $\Omega_{c}$, coherent drive on the $ac$ transition always suppresses the backward gain. In Fig.(\ref{CETLFig11}), we have plot the output pulse when the drive field $\Omega_{c}=12\gamma_{b}$. Our results show that the pulse in the forward direction is much broader than in the backward direction.
If we replace the coherent drive by an incoherent pump $\Phi $, which does not induce coherence, the gain becomes smaller when $\Phi $ increases (see Fig. \ref{Fig10}) for forward and backward direction and also the symmetry is restored.
\section{Conclusion}
In this paper we report the effect of coherence on the transient lasing. First we illustrated a possibility of having transient lasing without population inversion in $\Lambda -$scheme when spontaneous decay rate of the driving transition $\gamma _{c}$ is greater than those of the lasing transition $\gamma _{b}$. However, such condition is usually not satisfied for lasing at shorter wavelength as the spontaneous decay rate is proportional of the third power of the frequency. Having in mind improving performance of XUV and X-ray lasers with inversion by driving a longer wavelength optical transition, we consider $\Lambda $ and Cascade schemes with $\gamma _{b}\gg \gamma _{c}$. To show the effect of coherence we first optimize parameters of the model in the absence of the driving field, i.e., find the sample length for the fixed initial populations which yields the maximum output energy of the laser pulse. Then we drive the $a\leftrightarrow c$ transition with a coherent source $\Omega _{c}$ or an incoherent pump $\Phi $. We demonstrate that coherent drive can yield substantial enhancement (an order of magnitude) of the laser pulse energy for highly inverted medium than in the absence of the coherent drive [see Fig. 7(a)], while incoherent pump results in energy decrease (see Fig. 8). Contrary to the forward direction, where forward gain can be enhanced for certain choice of the drive Rabi frequency $\Omega_{c}$, coherent drive on the $ac$ transition always suppresses the backward gain (see Fig. 10). Thus, implementation of a coherent drive at optical frequency could be a useful tool for improving performance of lasers in XUV and X-ray regions.

\section{Acknowledgement}
We thank Szymon Suckewer and Yuri Rostovtsev for useful discussions and gratefully acknowledge support of this work by the National Science Foundation Grant EEC-0540832 (MIRTHE ERC), the Office of Naval Research, and the Robert A. Welch Foundation (A-1261). P. K. Jha also acknowledges Herman F. Heep and Minnie Belle Heep Texas A\&M University Endowed Fund held/administered by the Texas A\&M Foundation and Post-Doctoral Fellowship, Robert A. Welch Foundation.
 

\begin{thebibliography}{99}
\bibitem{J1} Here in this paper we have used the definition of Rabi frequency $\Omega=\wp\mathcal{E}/2\hbar$ viz. $\Omega_{b}=\wp_{ab}\mathcal{E}_{b}/2\hbar$
\bibitem{Comment1} The coherence for the forward $(+)$ and the backward $(-)$ propagating field are being treated as independent.
\bibitem{Comment2} We have neglected any contribution from $\Omega^{-}_{c}$ as nearly all the population in the level $a$ will be consumed during lasing in $ab$ transition under the condition $\gamma_{b} \gg \gamma_{c}$. Even if we include this small contribution from $\Omega^{-}_{c}$, the result shown in Fig \ref{CETLFig10} remains intact.
\end{thebibliography}
\end{document}